\documentclass[11pt]{article}
\usepackage{jhip}
\usepackage{amsmath,amssymb,amsfonts,amsbsy}

\usepackage{hyperref}

\newcommand{\nc}{\newcommand}
\nc{\rnc}{\renewcommand }
\nc{\tep}{\tilde{\epsilon}}
\rnc{\d}{\mathrm{d}}
\nc{\D}{\partial}
\nc{\bg}{\bar{g}}
\nc{\g}{\gamma}
\rnc{\o}{\omega}
\nc{\n}{{(n)}}
\rnc{\t}{\tau}
\nc{\ep}{\epsilon}
\nc{\K}{\mathcal{K}}
\nc{\cK}{\mathcal{K}}
\nc{\J}{\mathcal{J}}
\rnc{\H}{{\mathcal{H}}}
\nc{\lrarrow}{\leftrightarrow}
\nc{\nn}{\nonumber}
\nc{\tto}{\rightarrow}
\rnc{\[}{\begin{equation}}
\rnc{\]}{\end{equation}}
\nc{\bea}{\begin{eqnarray}}
\nc{\eea}{\end{eqnarray}}
\rnc{\(}{\left(}
\rnc{\)}{\right)}
\nc{\half}{\frac{1}{2}}

\nc{\p}{\partial}
\nc{\s}{\sigma}
\nc{\eps}{\epsilon}

\title{\mbox{}\\ \vspace{1cm} The relativistic fluid dual to vacuum Einstein gravity}

\author[a,b]{Geoffrey Comp\`ere,}
\author[b,d]{Paul McFadden,}
\author[a,b,c]{Kostas Skenderis}
\author[b,c]{and Marika Taylor}

\affiliation[a]{KdV Institute for Mathematics, }
\affiliation[b]{Institute for Theoretical Physics, }
\affiliation[c]{Gravitation and Astro-Particle Physics Amsterdam, \\[0.5ex] Science Park 904, 1090 GL Amsterdam, the Netherlands.}
\affiliation[d]{Perimeter Institute for Theoretical Physics, Waterloo ON N2L 2Y5, Canada.}

\emailAdd{gcompere@uva.nl}
\emailAdd{pmcfadden@perimeterinstitute.ca}
\emailAdd{K.Skenderis@uva.nl}
\emailAdd{M.Taylor@uva.nl}

\abstract{
We present a construction of a $(d+2)$-dimensional Ricci-flat metric
corresponding to a $(d+1)$-dimensional relativistic fluid, representing holographically
the hydrodynamic regime of a (putative) dual theory.  We show how to obtain the metric to arbitrarily high
order using a relativistic gradient expansion, and explicitly carry out the computation to second order.
The fluid has zero energy density in equilibrium, which implies incompressibility at first order in gradients, and its stress tensor (both at and away from equilibrium) satisfies a quadratic constraint, which determines its energy density away from
equilibrium. The entire dynamics to second order is encoded in one first order and six second order  transport coefficients, which we compute.  We classify entropy currents with non-negative divergence at second order in
relativistic gradients. We then verify that the entropy current obtained by pulling back to the fluid surface the area form at the null horizon indeed has a non-negative divergence.
We show that there are distinct near-horizon scaling limits that are equivalent either to the relativistic gradient expansion
we discuss here, or to the non-relativistic expansion associated with the Navier-Stokes equations discussed in previous works. The latter expansion may be recovered from the present relativistic expansion upon taking a specific non-relativistic limit.
}

\begin{document}
\maketitle

\section{Introduction and summary of results}

During the last year an interesting ``holographic'' connection between Ricci-flat metrics and fluids was uncovered:
in \cite{Bredberg} an approximate regular $(d+2)$-dimensional Ricci-flat metric corresponding to a solution of the incompressible non-relativistic Navier-Stokes equations in $(d+1)$ dimensions was presented, valid to leading non-trivial order in a non-relativistic hydrodynamic expansion, and in \cite{firstpaper}
we provided a systematic and unique construction of this metric to all orders. The construction has been extended to first non-trivial order to spherical horizons in vacuum gravity \cite{Bredberg:2011xw,Huang:2011he,Nakayama:2011bu}, to de Sitter horizons \cite{Anninos:2011zn} and to higher-derivative theories coupled to matter \cite{Chirco:2011ex,Niu:2011gu}.
Important earlier related works include the membrane paradigm \cite{Damour1,Damour2, Thorne} and the more recent construction of solutions of AdS gravity describing the hydrodynamic regime of CFTs
in the context of the AdS/CFT correspondence
\cite{Bhattacharyya:2008jc,Fouxon:2008tb,Bhattacharyya:2008kq,Eling:2009pb},
see also earlier results in \cite{Policastro:2001yc}.
Another instance of the fluid/gravity correspondence in asymptotically flat spacetimes can be found in the blackfold approach \cite{Emparan:2009cs,Emparan:2011hg}.
Further developments were reported in \cite{Cai:2011xv,Mei:2011gv,Lysov:2011xx,Kuperstein:2011fn,Brattan:2011my,Verschelde:2011wr,Eling:2011ct,Rodrigues:2011gg,Huang:2011kj,Marolf:2012dr}.

The existence of fluid solutions in gravity is expected/predicted by holography on general grounds. A generic feature of QFTs is the existence of a hydrodynamic description capturing the long-wavelength behaviour near to thermal equilibrium. One then expects to find the same feature on the dual gravitational side, i.e., there should exist a bulk solution corresponding to the thermal state, and nearby solutions corresponding to the hydrodynamic regime. Global solutions corresponding to near-equilibrium configurations should be well approximated by the solutions describing the hydrodynamic regime at sufficiently long distances and late times. This picture in indeed beautifully realised in the AdS/CFT correspondence, where the thermal state corresponds to a bulk black hole \cite{Witten1998}, and nearby solutions describing the
hydrodynamic regime, corresponding to solutions of relativistic conformal fluid mechanics, were constructed in \cite{Bhattacharyya:2008jc}. These solutions
were obtained by starting from the general equilibrium configuration, promoting the parameters characterising it
(temperature, relativistic velocities, etc.) to slowly
varying functions of spacetime, and then solving the bulk field equations iteratively in a derivative expansion. A further non-relativistic
limit leads to a correspondence between metrics of constant negative curvature and solutions of the incompressible non-relativistic Navier-Stokes equations
of the underlying conformal fluid \cite{Fouxon:2008tb,Bhattacharyya:2008kq}.

\begin{figure}[t]
\begin{center}
\includegraphics[width=4cm]{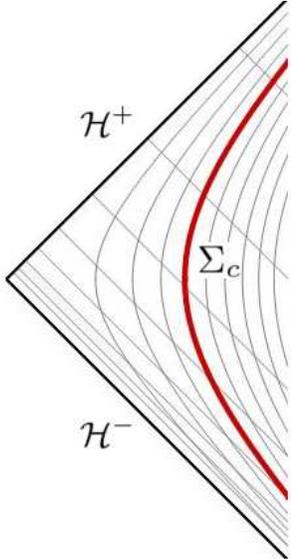}\hspace{1cm}
\begin{minipage}{0.5\textwidth}
\vspace{-6.5cm}
\caption{The analogue of the thermal state in our construction is Rindler spacetime, with past and future horizons $\H^-$ and $\H^+$ respectively.  The dual fluid lives on a constant acceleration surface $\Sigma_c$ with flat induced metric.  Lines of constant $\tau$ and constant $r$ in the coordinate system \eqref{Rindler} are shown in grey.} \label{fig:Rindler}
\end{minipage}
\end{center}
\end{figure}

In our construction the analogue of the thermal state is Rindler spacetime (see figure \ref{fig:Rindler}), and the first step in \cite{firstpaper}
was to obtain the general equilibrium
configuration. Nearby solutions describing the approach to equilibrium were then obtained by promoting the parameters appearing in the
equilibrium solution to slowing varying functions of spacetime. Finally, the bulk equations were solved iteratively by applying a non-relativistic hydrodynamic expansion, which we will call the $\ep$ expansion.
More precisely, the incompressible Navier-Stokes equations have a scaling symmetry (that, in particular, scales space and time non-relativistically),
and higher-derivative corrections to the Navier-Stokes equations are naturally organised according to their scaling. At third order in the $\ep$ expansion,
one finds a bulk metric
corresponding to solutions of the incompressible non-relativistic Navier-Stokes equations \cite{Bredberg}, while at higher orders the bulk metric corresponds
to solutions of the Navier-Stokes equations corrected by specific higher-derivative corrections \cite{firstpaper}.

In \cite{firstpaper} it was observed that all information could be recovered from a relativistic dissipative fluid by taking a non-relativistic limit. As explained in \cite{firstpaper}, the $\ep$-expansion never decreases the number of derivatives, but it may increase it, so the complete answer up to a given order in $\ep$ may be obtained starting from a relativistic dissipative stress tensor containing a sufficient number of dissipative terms but the converse is not in general true.
Indeed, the relativistic expansion is considerably more compact: we found that almost the entire information up to order $\ep^5$ is encoded in just one first order and four second order coefficients, while two further second order transport coefficients were undetermined at this order in the $\ep$ expansion.  In fact, only two terms in the non-relativistic stress tensor up to order $\ep^5$ were not recovered, but both of these required starting from a relativistic stress tensor of third order in gradients.

These results indicate that there is a manifestly relativistic construction, and one of the aims of this paper is to flesh this out:
instead of solving the bulk equations iteratively in the non-relativistic hydrodynamic expansion, we will solve them in a relativistic derivative expansion. This new expansion is significantly more powerful, and allows a treatment of the entropy currents which would be hard without going to very high order in the previous non-relativistic expansion.  We are also able to compute the previously undetermined coefficients in the second order expansion of the fluid stress tensor, as well as recovering our earlier results for the other coefficients. Thus the current treatment is the Ricci-flat analogue of the AdS treatment \cite{Bhattacharyya:2008jc}, while our previous construction is the analogue of \cite{Fouxon:2008tb,Bhattacharyya:2008kq}.

Let us now summarise the results for the stress tensor up to second order in gradients, which in the gauge where $T_{ab}u^b h^a_c = 0$ takes the form
\bea
T_{ab} = \rho u_a u_b + p h_{ab} + \Pi_{ab}^\perp, \qquad  \Pi_{ab}^\perp u^a = 0. \label{stress1}
\eea
One of the main results is that the fluid dual to vacuum Einstein gravity has zero equilibrium energy density
\bea \label{equilrho}
\rho_{eq} = 0\,.
\eea
Moreover, the stress tensor, including dissipative terms, satisfies the quadratic constraint,
\bea  \label{qua-con}
d T_{ab}T^{ab} = T^2.
\eea
This constraint determines $\rho$ as a function of $p$ and $\Pi_{ab}^\perp$ and, as such, it may be considered as a generalised equation of state.
When the relation is applied at equilibrium it leads to a quadratic equation with one of the two roots being (\ref{equilrho}).

The remaining freedom in defining the fluid variables is usually removed by redefining the energy density so that $T_{ab}u^a u^b = \rho_{eq}$. This so-called Landau gauge cannot be reached here since the equilibrium energy density is zero. Instead, we take the isotropic gauge by imposing that $\Pi_{ab}^\perp$ does not contain terms proportional to $h_{ab}$. A general fluid in flat spacetime and at first order in gradients is determined by two first order coefficients: the shear viscosity and bulk viscosity. A short computation shows, however, that when the equilibrium energy density is zero the conservation of the stress
tensor at leading order implies that the fluid is incompressible (to this order). The
bulk viscosity may then be replaced by another parameter $\zeta'$ which measures variations of the energy density at first order in gradients, $\rho^{(1)} =\zeta' D\ln p$. Up to second order in gradients, one needs 11 additional second order coefficients (in flat spacetime), namely
\begin{align}
\rho &= \zeta' D\ln p + \frac{1}{p}\left( d_1 \K_{ab}\K^{ab} + d_2 \Omega_{ab}\Omega^{ab} + d_3(D \ln p )^2 +d_4 DD \ln p +d_5 (D_\perp \ln p )^2 \right), \label{rho2a} \\
\Pi_{ab}^\perp &= -2\eta \K_{ab} + \frac{1}{p}\left( c_1 \K_a^c\K_{cb} + c_2 \K_{(a}^c\Omega_{|c|b)} + c_3 \Omega_a^{\,\,\,c}\Omega_{cb} +
c_4 h_a^ch_b^d\D_c\D_d \ln p \right. \nn\\
& \qquad\qquad\qquad\quad  \left. + c_5 \K_{ab}\,D\ln p + c_6 D^\perp_a \ln p \,D^\perp_b\ln p \right) \, .\label{stress2}
\end{align}
One of our main results in this paper is the computation of all the above coefficients for the fluid dual to vacuum Einstein gravity,
\bea
&& \zeta' = 0 ,\qquad d_1 = -2, \qquad d_2=d_3=d_4=d_5 = 0, \nn\\
&& \eta = 1,\qquad c_1=-2,\quad c_2 = c_3 = c_4 = c_5 = -c_6 = -4\, .\label{solc}
\eea
As noted earlier, for theories satisfying the constraint (\ref{qua-con}), the energy density (\ref{rho2a})
is not independent but rather follows from \eqref{qua-con},
so all in all the dual fluid is determined by one first order (the shear viscosity) and six second order coefficients (the $c$ coefficients).

A general feature of systems away from equilibrium is that they possess an entropy current with non-negative divergence. At equilibrium this current
should reduce to the conserved entropy current, $\J_{eq}^a= s_{eq} u^a$, where $s_{eq}$ the entropy density at equilibrium.  In the hydrodynamic regime,
the entropy current may differ from this expression by terms of higher order in gradients. Here we classify the possible entropy currents with non-negative
divergence for fluids with vanishing equilibrium energy density up to second order in gradients (in flat spacetime).
It turns out that first order gradients are not allowed and that there is a five-parameter family of allowed entropy currents depending on second order gradients:
\begin{align} \label{entr_curr}
\J^a &= s_{eq} u^a \Big( 1 +\frac{1}{p^2}(a_1 \K_{ab}\K^{ab}+a_2 \Omega_{ab}\Omega^{ab} -\frac{1}{2}(4 a_2 - 5 b_1 + 4 b_2 + b_3) (D_\perp \ln p)^2 )\Big) \nn\\
&+\frac{s_{eq}}{p^2}\left( b_1 h^{ac}\p_b \K^b_c +b_2 D_\perp^a D \ln p +b_3 \K^a_b D_\perp^b \ln p + (2 b_1 - 2 b_2 - b_3) \Omega^{ab}D_b^\perp \ln p  \right. \nn \\
& \qquad \qquad \left.+ (4 a_2 - 5 b_1 + 3 b_2 + b_3) D_\perp^a \ln p D\ln p \right).
\end{align}
A two-parameter subset of these entropy currents is in fact trivially conserved.

One may then ask what is the entropy current associated with the fluid dual to vacuum Einstein gravity. To compute this we have adapted the construction of \cite{bbbb}, defining the boundary entropy current by pulling back a suitable horizon quantity along certain null geodesics. This leads to an entropy current of the form (\ref{entr_curr}), i.e., with non-negative divergence, with coefficients
\bea
a_1=1, \qquad a_2 = \frac{1}{2}, \qquad b_1=-1, \qquad b_2 = -2, \qquad b_3=1.
\eea
Given the unconventional properties of the fluid dual to vacuum Einstein gravity, it is reassuring that
the entropy current indeed has non-negative divergence.

It was observed in \cite{Bredberg} that from the bulk perspective, the non-relativistic expansion could be expressed as a combination of a Weyl rescaling plus a particular near-horizon limit. A similar bulk interpretation also exists for the relativistic expansion we consider here.  As one takes the near-horizon limit, however, there are different quantities
that one keeps fixed in the two cases, so the two limits are distinct. Of course, as discussed earlier, one may always
take a further non-relativistic limit to go from the relativistic to the non-relativistic case.

This paper is organised as follows. In the next section, we present the relativistic construction of the
near-equilibrium solutions. Then, in section \ref{sec:sol}, we present the solution to second order in gradients,
and in section \ref{sec:entropy}, we discuss the classification of entropy currents to second order in gradients
and the holographic computation of the entropy current. In section \ref{sec:near}, we discuss the near-horizon limits
and we conclude in section \ref{sec:disc}.
 Finally, in the appendix we present a basis for scalars, vectors and tensors,
up to the order required for the hydrodynamic analysis.

{\bf Note added: } During the completion of this paper we were informed about the forthcoming publication \cite{Elingtoappear}
which has significant overlap with the material presented here.  The results of \cite{Elingtoappear} are in agreement with those presented here.

\section{Relativistic construction of near-equilibrium solutions}

\subsection{Seed metric}
\label{sec:seed}

We start with Minkowski spacetime in Rindler coordinates,
\[
\label{Rindler}
\d s^2 = -r \d\t^2+2\d\t\d r+\d x_i\d x^i.
\]
The metric $\gamma_{ab}$ on the surface $\Sigma_c$ defined by $r=r_c$ is
\[
\gamma_{ab}\d x^a\d x^b = -r_c \d\t^2+\d x_i\d x^i,
\]
where the coordinates $x^a = (\t, x^i)$.
To obtain the zeroth order seed metric we perform the following changes of coordinate:
first, we send
\[
r \tto r+1/p^{2}-r_c, \qquad \t \tto \sqrt{r_c} p \t,
\]
taking the metric to
\[
\d s^2 = 2p (\sqrt{r_c}\d \t) \d r -p^2(r-r_c) (\sqrt{r_c}\d\t)^2+\gamma_{ab}\d x^a \d x^b,
\]
and second, we perform the boost
\[
\sqrt{r_c} \t \tto - u_a x^a, \qquad
x^i \tto x^i-u^i\sqrt{r_c}\t+ (1+\gamma)^{-1}u^iu_jx^j
\]
where
\[
\label{relueq}
u_a =\frac{1}{\sqrt{r_c-v^2}}(-r_c, v_i), \qquad \gamma = (1-v^2/r_c)^{-1/2}.
\]
Since this boost preserves $\gamma_{ab}\d x^a\d x^b$, we arrive at the metric
\[
\label{seed_metric}
\d s^2 = -2p u_a \d x^a \d r + [\gamma_{ab}-p^2(r-r_c)u_au_b]\d x^a \d x^b.
\]
A simple calculation reveals that, with this metric, the Brown-York stress tensor on $\Sigma_c$ takes the form of a perfect fluid in equilibrium:
\[
T_{ab} = p h_{ab}, \qquad h_{ab}=\gamma_{ab}+u_a u_b \, .
\]
The position of the Rindler horizon is now $r_\H = r_c-1/p^2$, as may be seen by writing the metric in the form
\[
\d s^2 = -2 p u_a \d x^a \d r + [h_{ab}-p^2(r-r_\H)u_au_b]\d x^a\d x^b.
\]

In the remainder of this paper we will set $r_c\tto 1$ (taking $\gamma_{ab}\tto \eta_{ab}$).
This may be accomplished without loss of generality via the scaling
\[ \label{rescale}
(r,\t,x_i,r_c,p,v_i) \tto (\lambda^2 r, \t, \lambda x_i, \lambda^2 r_c, \lambda^{-1} p, \lambda v_i).
\]
Acting on the equilbrium metric this sends $\d s^2 \tto \lambda^2 \d s^2$, after which the constant overall conformal factor may be dropped.
One may similarly restore $r_c$ at any point by reversing this procedure.

\subsection{Integration scheme}

We start with the zeroth order seed metric \eqref{seed_metric} in the form
\[
 \d s^2 = -2p u_a \d x^a \d r+[\eta_{ab}+(1-\theta)u_au_b]\d x^a\d x^b, \qquad \theta = 1+p^2(r-1)
\]
where we have scaled $r_c$ to unity and introduced the quantity $\theta$ for our later convenience. The position of the horizon is now $r_\H = 1-1/p^2$. Note also that the relativistic fluid velocity is normalised such that $\eta^{ab}u_au_b=-1$.
In this metric, the velocity and pressure $u_a$ and $p$ should now be regarded as functions of $x^a=(\tau, x^i)$.
The inverse of this metric is
\[
 g^{rr}=r-r_\H,\qquad g^{ra}=\frac{1}{p}u^a,\qquad g^{ab}=h^{ab},
\]
where we define
\[
h^{ab} \equiv \eta^{ab}+u^au^b, \qquad u^a\equiv \eta^{ab}u_b.
\]

Weighting derivatives such that $\p_r \sim 1$ and $\p_a \sim \tep$,
adding a piece $g^\n_{\mu\nu}$ to the metric at order $\tep^n$ engenders a change in the Ricci tensor
\begin{align}
\delta R^\n_{rr} &= -\half h^{ab}\p_r^2 g^\n_{ab}, \nn\\[1ex]
\delta R^\n_{ra} &=\frac{1}{2p}u^b\p_r^2 g^\n_{ab}-\frac{1}{4}u_a\Big[\frac{1}{p}\,\p_r(\theta  g^\n_{rr})+2u^b\p_r  g^\n_{br}-ph^{bc}\p_r  g^\n_{bc}\Big], \nn\\[1ex]
\delta R^\n_{ab} &= -\frac{1}{2p^2}\p_r\big(\theta \p_r g^\n_{ab}\big)- u^cu_{(a}\p_r g^\n_{b)c} \nn\\[1ex]&\quad
-\frac{1}{4}u_au_b \left[ \frac{\theta}{p^2}\,\p_r \big(\theta g^\n_{rr}\big)+\frac{2\theta}{p}u^c\p_r g^\n_{cr}+(2u^cu^d-\theta h^{cd})\p_r g^\n_{cd}\right].
\end{align}
A convenient gauge choice is $g^\n_{r\mu}=0$ for $n\ge 1$.  This choice eliminates metric components for which we do not have natural boundary conditions, and moreover ensures that worldlines of constant $x^a$ are bulk null geodesic to all order (see section \ref{subsec:entropydefn}).
The linearised Ricci tensor tensor is then
\begin{align}
\label{lin_Ricci}
\delta R^\n_{rr} &= -\half h^{ab}\p_r^2  g^\n_{ab}, \nn\\[1ex]
\delta R^\n_{ra} &=\frac{1}{2p}u^b\p_r^2  g^\n_{ab}+\frac{p}{4}u_ah^{bc}\p_r g^\n_{bc}, \nn\\[1ex]
\delta R^\n_{ab} &= -\frac{1}{2p^2}\,\p_r \big(\theta\p_r g^\n_{ab}\big) - u^cu_{(a}\p_r g^\n_{b)c}
-\frac{1}{4}u_au_b \left[(2u^cu^d-\theta h^{cd})\p_r g^\n_{cd}\right].
\end{align}
Setting $0=\delta R^\n_{\mu\nu}+\hat{R}^\n_{\mu\nu}$, we obtain the integrability conditions
\[
 0=\p_r\left(h^{ab}\hat{R}^\n_{ab}-\frac{\theta}{p^2}\,\hat{R}^\n_{rr}\right)-\hat{R}^\n_{rr}, \qquad
0 = \theta\hat{R}^\n_{ra} + p u^b\hat{R}^\n_{ab}.
\]
The first of these is the $r$-component of the Bianchi identity at order $\tep^n$; while for the second, the $a$-component of the Bianchi identity enforces
\[
  \theta\hat{R}^\n_{ra} + p u^b\hat{R}^\n_{ab} = f_a^\n(x).
\]
Evaluating the Gauss-Codazzi identity on $\Sigma_c$ at order $\tep^n$, we find
\[
\label{Gauss_Codacci}
\nabla^bT_{ab}\big|^\n_{\Sigma_c} = [2\nabla^b(K\g_{ab}-K_{ab})]^\n = [-2R_{a\mu}N^\mu]^\n =  -\frac{2}{p}f_a^\n(x).
\]
Thus, conservation of the Brown-York stress tensor at each order ensures that one may integrate the bulk equations.

The radial Hamiltonian constraint evaluated on $\Sigma_c$ is given by
\bea \label{Ham_con}
K^2 - K_{ab} K^{ab} =0,
\eea
which upon using the definition of the Brown-York stress tensor,
$T_{ab} = \frac{1}{8 \pi G} (K \gamma_{ab} - K_{ab})$, becomes the
quadratic constraint given earlier in \eqref{qua-con}. As was discussed
in \cite{firstpaper} and reviewed in the introduction,
this constraint plays the role of a generalised equation of state:
given $p$ and $\Pi^\perp_{ab}$ it determines $\rho$.

Note that the radial Hamiltonian constraint in AdS plays exactly the same
role, i.e., it determines the equation of state. Indeed, in the presence of
a cosmological constant the radial Hamiltonian constraint (in Fefferman-Graham gauge) 
becomes
\bea \label{Ham_con_AdS}
K^2 - K_{ab} K^{ab} = d(d-1),
\eea
where have set the AdS radius to unity and assumed a flat boundary metric.
Expanding about conformal infinity, the extrinsic curvature is given by \cite{Papadimitriou:2004ap}
\bea
K_{ab} = \eta_{ab} + K_{(d)ab} + \ldots,
\eea
where the subscript indicates dilatation weight and the dots represent higher order
terms that do not contribute when we evaluate the constraint at conformal
infinity. Inserting the holographic stress tensor
\cite{Papadimitriou:2004ap}
\bea
T_{ab} = 2 (K_{(d)}\eta_{ab} - K_{(d)ab}),
\eea
into the AdS Hamiltonian constraint (\ref{Ham_con_AdS}) then yields
\bea \label{traceT}
T^a_a=0,
\eea
which implies the equation of state of a conformal fluid.\footnote{ 
If we instead consider a general
boundary metric then the r.h.s.~of (\ref{traceT}) contains
the holographic Weyl anomaly.
}

Returning to the case of the Rindler fluid, a particular integral of \eqref{lin_Ricci} is
\[
 \tilde{g}^\n_{ab} = \tilde{\alpha}^\n u_a u_b +2\tilde{\beta}^\n_{(a}u_{b)} + \tilde{\gamma}^\n_{ab}, \qquad u^a\tilde{\beta}^\n_a=0, \qquad u^a\tilde{\gamma}^\n_{ab}=0,
\]
where
\begin{align}
\tilde{\alpha}^\n &= c_1^\n(x) +(1-r)c_2^\n(x) +2p^2 \int^{1}_r\d r'\int_{r'}^{1}\d r'' (h^{cd}\hat{R}^\n_{cd}-\half\hat{R}^\n) ,\nn\\[1ex]
\tilde{\beta}^\n_a &= c_{3a}^\n(x) +(1-r)c_{4a}(x) + 2p\int^1_r\d r'\int^1_{r'} \d r''  h^b_a\hat{R}^\n_{br}, \nn\\[1ex]
\tilde{\gamma}^\n_{ab} &= c_{5ab}^\n(x) +  c_{6ab}^\n(x) \ln\theta - 2p^2 \int^1_{r} \d r' \frac{1}{\theta}\int^{r'}_{r_*}\d r'' h_a^ch_b^d\hat{R}^\n_{cd},
\end{align}
where $\hat{R}^\n=g^{(0)\mu\nu}\hat{R}^\n_{\mu\nu}$.
Note that to satisfy the $rr$ equation, we must have that
\[
  2\hat{R}^\n_{rr}= h^{ab}\p_r^2\tilde{\gamma}^\n_{ab} =-\frac{p^4}{\theta^2}h^{ab}c_{6ab}^\n +\frac{2p^2}{\theta}h^{ab}\hat{R}^\n_{ab} -\frac{2p^4}{\theta^2}\int^r_{r_*}\d r' h^{ab}\hat{R}^\n_{ab}.
\]
Using the first integrability condition in the form
\[
 \p_r(h^{ab}\hat{R}^\n_{ab}) = \frac{1}{\theta}\p_r\left(\frac{\theta^2}{p^2}\hat{R}^\n_{rr}\right),
\]
one can show that
\[
 2\hat{R}^\n_{rr} = -\frac{2p^2}{\theta^2}\Big[\theta h^{ab}\hat{R}^\n_{ab}-\frac{\theta^2}{p^2}\hat{R}^\n_{rr}\Big]_* +\frac{2p^2}{\theta}h^{ab}\hat{R}^\n_{ab} -\frac{2p^4}{\theta^2}\int^r_{r_*}\d r' h^{ab}\hat{R}^\n_{ab},
\]
where $[\ldots]_*$ indicates evaluating at $r=r_*$.  One then obtains an equation for $r_*$, namely
\[
 h^{ab}c^\n_{6ab} = \frac{2}{p^2}\Big[\theta h^{ab}\hat{R}^\n_{ab}-\frac{\theta^2}{p^2}\hat{R}^\n_{rr}\Big]_*.
\]
Later, we will choose to set the coefficients of all logarithmic terms to zero.  For the trace component above, this entails setting the lower limit of integration such that $r_*=r_\H \equiv 1-1/p^2$, whereupon $\theta_*=0$.

Allowing now for gauge transformations $\xi^{\n\mu}$ at order $\tep^n$, as well as re-definitions $\delta u^{\n a}(x)$ and $\delta p^\n(x)$ of the fluid velocity and pressure, the solution above generalises to
\begin{align}
 g^\n_{rr}&=-2pu_a\p_r\xi^{\n a},\\
g^\n_{ra} &= -u_a[p\p_r\xi^{\n r}-(1-\theta)u_b\p_r \xi^{\n b}+\delta p^\n] +\eta_{ab}\p_r\xi^{\n b}-p\delta u_a^\n, \\
g^\n_{ab}&= \tilde{g}^\n_{ab} -u_au_b[p^2\xi^{\n r}+2p\delta p^\n(r-1)]+2(1-\theta)u_{(a}\delta u^\n_{b)}.
\end{align}
To impose the gauge choice $g^\n_{r\mu}=0$, we must then set
\[
 \xi^{\n r} = (1-r)\frac{\delta p^\n}{p}+\tilde{\xi}^{\n r}(x), \qquad \xi^{\n a}=\xi^\n u^a + \xi^{\n a}_\perp,
\]
where $u_a\xi^{\n a}_\perp=0$ and
\[
 \xi^\n=\tilde\xi^\n(x),\qquad \xi_\perp^{\n a}=-(1-r)p\delta u^{\n a}+\tilde{\xi}_\perp^{\n a}(x).
\]
The remaining metric components then take the form:
\begin{align}
\alpha^\n &= c_1^\n-p^2\tilde{\xi}^{\n r} +(1-r)(c_2^\n+p\delta p^\n) +2p^2 \int^{1}_r\d r'\int_{r'}^{1}\d r'' (h^{cd}\hat{R}^\n_{cd}-\half\hat{R}^\n) ,\nn\\[1ex]
\beta^\n_a &= c_{3a}^\n +(1-r)(c_{4a}+p^2\delta u^\n_a) + 2p\int^1_r\d r'\int^1_{r'} \d r''  h^b_a\hat{R}^\n_{br}, \nn\\[1ex]
\gamma^\n_{ab} &= c_{5ab}^\n +  c_{6ab}^\n \ln\theta - 2 \int^1_{r} \d r' \frac{1}{r'-r_\H}\int^{r'}_{r_*}\d r'' h_a^ch_b^d\hat{R}^\n_{cd}.
\end{align}
Imposing the boundary condition $g^\n_{ab}=0$ for $n\ge 1$ on $\Sigma_c$ then fixes
\[
 c_1^\n= p^2\tilde{\xi}^{\n r}, \qquad c_{3a}^\n=0,\qquad c_{5ab}^\n=0.
\]
Moreover, for regularity on the future horizon $r=r_\H$, we must set
\[
 c_{6ab}^\n=0, \qquad r_*=r_\H.
\]

In summary then, our integration scheme is the following.
The $rr$ and $ra$ metric components are given to all orders by
\[
g_{rr}=0,\qquad g_{ra}=-pu_a,
\]
while the $ab$ metric components may be decomposed in fluid variables as
\[
g^\n_{ab} = \alpha^\n u_a u_b +2\beta^\n_{(a}u_{b)} + \gamma^\n_{ab}.
\]
Beginning with the seed metric $g^{(0)}_{ab} =-\theta u_au_b+h_{ab}$, the metric at all subsequent orders is then given by
\begin{align}
\label{soln_algorithm}
\alpha^\n &= (1-r)F^\n(x) +2p^2 \int^{1}_r\d r'\int_{r'}^{1}\d r'' (h^{cd}\hat{R}^\n_{cd}-\half\hat{R}^\n) ,\nn\\[1ex]
\beta^\n_a &= (1-r)F_a^\n(x) + 2p\int^1_r\d r'\int^1_{r'} \d r''  h^b_a\hat{R}^\n_{br}, \nn\\[1ex]
\gamma^\n_{ab} &=  - 2 \int^1_{r} \d r' \frac{1}{r' - r_\H}\int^{r'}_{r_\H}\d r'' h_a^ch_b^d\hat{R}^\n_{cd}.
\end{align}
Here, the arbitrary functions
\[
 F^\n(x) = c_2^\n(x)+p\delta p^\n(x), \qquad  F_a^\n(x) = c_{4a}(x)+p^2\delta u^\n_a(x),
\]
encode the choice of gauge for the dual fluid, and will be fixed as we discuss in the following section.  Note also that $F_a^\n$ is transverse: $u^aF_a^\n=0$.

\subsection{The Brown-York stress tensor}

The variation in the extrinsic curvature of $\Sigma_c$ at order $\tep^n$ due to $g_{ab}^\n$ is
\[
\delta K^{(n)}_{ab}\big|_{\Sigma_c} = \frac{1}{2}\pounds_{N} g_{ab}^\n = \frac{1}{2}N^r\D_r g_{ab}^\n = \frac{1}{2p}\D_r g_{ab}^\n.
\]
(Note here that the normal $N^\mu|_{\Sigma_c}=p^{-1}\delta^\mu_r+\delta^\mu_a u^a$ to all orders, since the bulk metric at $\Sigma_c$ is effectively fixed.)
Evaluating this explicitly, we find
\begin{align}
\delta K^\n_{ab}\big|_{\Sigma_c} &= -\frac{1}{2p}F^\n u_au_b-\frac{1}{p}u_{(a}F^\n_{b)}+p\int^1_{r_\H}\d r' h^c_ah_b^d\hat{R}^\n_{cd}.
\end{align}
The variation in the Brown-York stress tensor due to $g_{ab}^\n$ is thus
\begin{align}
\label{deltaT}
 \delta T_{ab}^\n\big|_{\Sigma_c} &= 2(\eta_{ab}\delta K^\n-\delta K_{ab}^\n) \nn\\
&=\frac{1}{p}F^\n h_{ab}+\frac{2}{p}u_{(a}F^\n_{b)}+2p\int^1_{r_\H}\d r' (\eta_{ab}h^{cd}-h_a^c h_b^d)\hat{R}^\n_{cd}.
\end{align}
The complete Brown-York stress tensor on $\Sigma_c$ at order $\tep^n$ is then
\[
 T^{(n)}_{ab}\big|_{\Sigma_c} = \delta T_{ab}^\n + \hat{T}_{ab}^\n,
\]
where $\hat{T}_{ab}^\n$ represents the contribution at order $\tep^n$ due to the metric up to order $\tep^{n-1}$.

\subsection{Gauge choices for fluid}
\label{subsec:fluid_gauge}

We will define the relativistic fluid velocity $u^a$ such that
\[
0= h_a^bT_{bc}u^c,
\]
which then uniquely fixes $F^\n_a$:
\[
F^\n_a = p h_a^b\hat{T}^\n_{bc}u^c.
\]

To fix $F^\n$, we impose that there are no corrections to the pressure, i.e., that the coefficient of $h_{ab}$ in $T_{ab}$ is fixed to be exactly $p$.
From \eqref{deltaT}, we see that $F^\n$ is then determined uniquely.

\section{Solution} \label{sec:sol}

In the previous section, we saw how to systematically construct the near-equilibrium solution in terms of a relativistic gradient expansion starting from the seed solution.
We saw moreover that the solution is unique once we impose the bulk gauge conditions, the gauge conditions on the fluid stress tensor, and regularity at each order in the expansion.
In the present section, we will now explicitly compute this solution to second order.

\subsection{First order}

Computing the Ricci curvature of the seed metric, we obtain
\[
\hat{R}^{(1)}=0,\qquad \hat{R}^{(1)}_{r\mu}=0,\qquad \hat{R}^{(1)}_{ab} = \big(Dp+\frac{p}{2} \p_cu^c\big)u_au_b+pu_{(a}a_{b)}+u_{(a}\p_{b)}p,
\]
where $D\equiv u^a\p_a$, $D^\perp_a\equiv h_a^b\p_b$ and the acceleration $a_c=Du_c$.
Since $h^{ab}\hat{R}^{(1)}_{ab}$ vanishes, the integration step is trivial and we have
\[
g^{(1)}_{ab} = (1-r)\big[F^{(1)}u_au_b+2F^{(1)}_{(a}u_{b)}\big].
\]
Evaluating the Brown-York stress tensor on $\Sigma_c$, we find
\[
 T_{ab}\big|_{\Sigma_c} =\Big(p+2D\ln p+\frac{1}{p}F^{(1)}\Big)h_{ab}+2u_{(a}\Big(2a_{b)}+\frac{1}{p}F^{(1)}_{b)}\Big) -2\mathcal{K}_{ab},
\]
where we write the fluid shear and vorticity
\[
 \mathcal{K}_{ab}=h_a^ch_b^d\p_{(c}u_{d)},\qquad  \Omega_{ab}=h_a^ch_b^d\p_{[c}u_{d]},
\]
so that
\[
 \p_au_b = \mathcal{K}_{ab}+\Omega_{ab}-u_a a_b.
\]
Note here that conservation of the stress tensor at zeroth order yields the conditions
\[
\p_cu^c=O(\p^2),\qquad  a_c + D_c^\perp \ln p =  O(\p^2),\label{eq:zeroth}
\]
and that we used both of these to simplify the form of the stress tensor at first order.

Our gauge condition for the pressure immediately sets
\[
 F^{(1)}=-2 D p,
\]
while that for the fluid velocity at first order reads
\[
0= h_a^bT_{bc}u^c = -\frac{1}{p}F^{(1)}_a - 2a_a \quad \Rightarrow \quad F^{(1)}_a = -2p a_a.
\]
In conclusion then,
\[
 g^{(1)}_{ab}=2(r-1)( u_au_bDp+2p a_{(a}u_{b)})\label{g1}
\]
while
\[
 T_{ab}\big|_{\Sigma_c}= ph_{ab} -2\mathcal{K}_{ab}+O(\p^2),\label{stress1bis}
\]
from which we may read off the viscosity $\eta=1$. We also deduce that the first order coefficient $\zeta'=0$, where $\zeta'$ is defined in \eqref{stress1}-\eqref{rho2a}.

\subsection{Second order}

The fluid equations of motion at second order may be directly obtained from conservation of the first order stress tensor \eqref{stress1} with $\zeta' = 0$, $\eta = 1$ as
\[
\p_b u^b - \frac{2}{p}\mathcal K_{bc}\mathcal K^{bc} = O(\p^3)\, , \qquad
a_a + D_a^\perp \ln p -\frac{2}{p}h_a^c \p_b \mathcal K_c^b = O(\p^3)\, .\label{eq:f1}
\]
Next, from the first order metric \eqref{g1}, one may obtain the second order piece of the Ricci tensor $\hat R^{(2)}_{\mu\nu}$. The computation is straightforward but laborious. Using \eqref{rel1} - \eqref{rel4} and the fluid equations \eqref{eq:f1}, the result may be expressed in the basis of tensors given in the appendix.  We find
\begin{align}
\hat R_{rr}^{(2)} &= -p^2 \Omega^{ab}\Omega_{ba} \, \nn,\\
\hat R_{ra}^{(2)}&= p u_a \Big(  ( D_\perp \ln p)^2 +\K_{bc}\K^{bc}   +p^2 (r-1)  \Omega_{bc}\Omega^{bc}   \Big) +p h_a^c \p_b \K^b_{\; \,c}   +p (\K_{ab}+\Omega_{ab})D_\perp^b\ln p   \, \nn,\\
\hat R_{ab}^{(2)} &= u_a u_b \theta(r) \Big(   (D_\perp \ln p )^2  +\K_{cd}\K^{cd} +p^2 (r-1) \Omega_{cd}\Omega^{cd} \Big) +2u_{(a} \theta(r) \Big(  (\mathcal K_{b) c}+\Omega_{b) c})D_\perp^c \ln p  \nn \\
&\quad  + h_{b)}^c \p_d \K^d_{\; \, c)}) \Big)  + 2\K_{ab}D\ln p + 2h_a^c h_b^d \p_c \p_d \ln p -2  D_a^\perp \ln p D_b^\perp \ln p+\K_a^c \K_{cb}+ \Omega_{a}^{\;\, c}\Omega_{cb} \nn\\
&\quad + 2 \mathcal K_{c(a}\Omega^c_{\; \, b)} +2p^2(r-1)\Omega_{c(a}\Omega^c_{\; \, b)}\, .\label{R2_3}
\end{align}
(Note the equations $N^\mu R_{\mu a} = O(\p^3)$ are satisfied, as required by \eqref{Gauss_Codacci}.)

As noted earlier, the stress tensor at second order $T_{ab}^{(2)}$ evaluated on $\Sigma_c$ is the sum of two pieces:
the piece $\delta T_{ab}^{(2)}$ resulting from the second order contribution to the metric $g^{(2)}_{ab}$,
and the piece $\hat{T}_{ab}^{(2)}$ resulting from evaluating the stress tensor to second order using the metric up to first order.
Computing the stress tensor at first order without using the first order fluid equations of motion, we obtain
\bea
T_{ab} = p h_{ab} - 2\mathcal K_{ab}+2\mathcal K (h_{ab}-u_a u_b)-2u_{(a}\Big( a_{b)}+D^\perp_{b)}\ln p \Big)+O(\p^2)\, .
\eea
Using the equations of motion at second order \eqref{eq:f1}, we deduce that
\bea
\hat T_{ab}^{(2)} = \frac{4}{p}\K_{cd}\K^{cd}(h_{ab}-u_a u_b)-\frac{4}{p}u_{(a} \Big( h^c_b \p_d \K^d_{\; \, c} \Big)\, .
\eea
Next, inserting the second order Ricci tensor \eqref{R2_3} into \eqref{deltaT}, and making use of the substitutions listed in the appendix, we find
\begin{align}
\delta T^{(2)}_{ab} &= h_{ab} \left( \frac{1}{p}F^{(2)} - \frac{2}{p}(\mathcal K_{ab}\mathcal K^{ab}) \right)+\frac{2}{p} u_a u_b \Big( \mathcal K_{ab} \mathcal K^{ab}  \Big)  + \frac{2}{p}u_{(a}F_{b)}^{(2)}\nn \\
&\quad -\frac{2}{p}\Big( h_{(a}^d D_{b)}^\perp D^\perp_d \ln p - D^\perp_a \ln p D^\perp_b \ln p -h_a^c h^d_b D \mathcal K _{cd}+2 \mathcal K_{c(a}\Omega^c_{\; \, b)}-\Omega_{c(a}\Omega^c_{\; \, b)} \Big)\, .
\end{align}
Combining these two contributions to the second order stress tensor, our fluid gauge conditions in section \ref{subsec:fluid_gauge} are met when
\bea
F^{(2)} = -2 \mathcal K_{ab}\mathcal K^{ab},\qquad F^{(2)}_a = 2 h_a^c \p_d \mathcal K^d_c\, .
\eea
The fluid stress tensor then takes the expected form \eqref{stress1}, with $\rho$ and $\Pi^\perp_{ab}$ as given in \eqref{rho2a}-\eqref{stress2} and with the advertised second order coefficients \eqref{solc}.

As a consistency check, we note that our expression for $\rho^{(2)}$,
\bea
\rho^{(2)} = -\frac{2}{p}\mathcal K_{ab}\mathcal K^{ab}\, ,\label{rho2}
\eea
as well as the coefficients $c_1$, $c_2$, $c_3$ and $c_4$ listed in \eqref{solc}
coincide with our earlier results obtained in \cite{firstpaper} using the non-relativistic $\eps$-expansion to order $O(\eps^6)$. The two remaining coefficients, $c_5$ and $c_6$, that were not determined in the analysis of \cite{firstpaper} have moreover now been computed.

Recalling that the second order metric takes the form
\bea
g_{ab}^{(2)} = \alpha^{(2)}u_a u_b + 2\beta^{(2)}_{(a}u_{b)}+\gamma^{(2)}_{ab}\, ,
\eea
using the solution algorithm \eqref{soln_algorithm} with our result \eqref{R2_3} as input, we obtain
\begin{align}
\label{alpha2a}
\alpha^{(2)}&= 2(r-1)\mathcal K_{ab}\mathcal K^{ab}+p^2 (\frac{1}{2}\mathcal K_{ab}\mathcal K^{ab}+ (D_\perp \ln p )^2 )(r-1)^2 +\frac{p^4}{2}(r-1)^3 \Omega_{ab}\Omega^{ab} \, ,\nn\\
\beta^{(2)}_a &= -2 (r-1)h_a^c \p_d \mathcal K^d_{\;c} +p^2(r-1)^2 (h_a^b \p_c \mathcal K^c_{\; b}+(\mathcal K_{ab}+\Omega_{ab})D^b_\perp \ln p ) \, ,\nn\\
\gamma^{(2)}_{ab}&= 2(r-1)\Big( 2 h_a^c h_b^d \p_c \p_d \ln p +2 \mathcal K_{ab} D\ln p-2D^\perp_a \ln p D^\perp_b \ln p+\mathcal K_a^{\; c}\mathcal K_{cb} \nn \\
&\qquad\qquad\qquad +2 \Omega_a^{\; \, c}\Omega_{cb}+2 \mathcal K_{c(a}\Omega^c_{\; b)}\Big)+p^2(r-1)^2 \Omega_{ca}\Omega^{c}_{\; \, b}\, .
\end{align}
As a check of this result, one may proceed to expand the metric we have just calculated to $O(\eps^6)$ in the non-relativistic $\eps$-expansion.  One should then recover all terms of up to second order in gradients in the metric of \cite{firstpaper}, modulo a small complication which is that the metric in \cite{firstpaper} was computed in a slightly different radial gauge. (The choice of fluid gauge in \cite{firstpaper} is also slightly different.)
To get round this, we simply repeated the analysis of \cite{firstpaper} with our current gauge choices.  Comparing with the $\eps$-expansion of $g_{ab}^{(2)}$, we then found that the two results are indeed consistent.

In summary, the complete metric up to second order is given by
\[
g_{rr} = 0,\qquad g_{ra} = - p u_a, \qquad 
g_{ab} = \alpha u_a u_b + 2\beta_{(a}u_{b)}+\gamma_{ab} \, , \label{secondorderg}
\]
where
\bea
\alpha &=& - 1 -p^2 (r-1)+2 p D\ln p (r-1) +\alpha^{(2)} \, ,\nn\\
\beta_a &=& 0 + 2 p a_a (r-1) + \beta^{(2)}_a \, ,\nn\\
\gamma_{ab} &=& h_{ab} + 0 + \gamma_{ab}^{(2)}\, .
\eea
The inverse metric up to second order is
\bea
\label{invg2}
g^{rr} &=& \frac{1}{p^2}\Big( - \alpha + h^{ab} \beta_a \beta_b \Big) \, ,\nn\\
g^{ra} &=& \frac{1}{p}\Big( u^a +h^{ab} \beta_b \Big) \,, \nn\\
g^{ab} &=& h^{ab} - h^{ac} h^{bd} \gamma^{(2)}_{cd}\, .
\eea

\subsection{Fluid divergence at third order}

Given the stress tensor at second order, it is straightforward to obtain the fluid equations of motion at third order.
We will need to make use of the third order fluid continuity equation $u^a \p^b T_{ab} = 0$ in our forthcoming discussion of the entropy current.
Let us then derive this equation for a general fluid stress tensor of the form \eqref{stress1} in combination with \eqref{rho2a}-\eqref{stress2}. Using the basis of tensors given in the appendix, we obtain
\bea
\D_a u^a &=& \frac{1}{p}\left[ 2 \eta \K_{ab} \K^{ab}-\zeta D^2\ln p \right] + \frac{1}{p^2} \left[
 (- c_4 + 2d_1) \K^{ab} \D_a \D_b \ln p  + (- c_1 +2 d_1 ) \K^{ab} \K_a^c \K_{bc} \right. \nn \\
 && \left.
 + (- c_3 +2d_1 -4d_2 )  \K^{ab} \Omega_a{}^c \Omega_{bc} + 3d_2 \Omega_{ab}\Omega^{ab}D\ln p
+(-c_6-2d_1) \K^{ab} D^\perp_a \ln p D^\perp_b \ln p \right. \nn \\
&&\left. +(c_4-c_5+d_1-2\eta \zeta )D\ln p \K_{ab}\K^{ab} -d_4 D^3 \ln p +(-2d_3 +d_4 +\zeta^2)D^2\ln p D\ln p \right. \nn\\
&& \left.
+d_3 (D\ln p )^2+d_5 (D_\perp \ln p )^2 D\ln p -2d_5 DD_\perp^a \ln p D^\perp_a\ln p
\right] +O(\p^4) \, .\label{generaldivv}
\eea
Interestingly, upon inserting the coefficients \eqref{solc} this equation simplifies to
\bea
\D_a u^a &=& \frac{2}{p} \K_{ab} \K^{ab} - \frac{2}{p^2} (\K^{ab} \K_a^c \K_{bc} + D \ln p  \K_{ab} \K^{ab}) +O(\p^4) \nn\\
&=& \frac{2}{p} \K_{ab} \K^{ac} \left(\delta_c^b  - \frac{1}{p}\K^b_c -\frac{D\ln p }{p}\delta_c^b \right)+O(\p^4)\, .
\eea
In particular, the right-hand side is non-negative since the third order terms are small corrections to the non-negative second order term.  This is a special property of the fluid dual to vacuum gravity:
in the general case, the divergence might be negative for flows where $\K_{ab}$ is small and other terms in the expression \eqref{generaldivv} dominate.

\section{Entropy current} \label{sec:entropy}

\subsection{General fluid entropy current}
\label{sec:entropys1}

A general entropy current is by definition a current which has non-negative divergence given the fluid equations of motion. We will derive in this section the constraints on the form of the entropy current at first and second order in the derivative expansion using solely the fluid equations of motion \eqref{eq:f1}.
In particular, our considerations do not depend on the second order part of the bulk metric.

Using dimensional analysis and the classification of first order scalars and vectors given in the appendix, the general entropy current at first order has the form
\bea
\J^a = s_{eq} \left( u^a + \frac{\alpha}{p} u^a D\ln p +\frac{\beta}{p} D^a_\perp \ln p  \right) +O(\p^2)\, .
\eea
Here, we normalised the current using the equilibrium entropy density $s_{eq}=1/(4G)$. Using the relationships in the appendix, the divergence of this entropy current is
\[
\frac{1}{s_{eq}}\p_a \J^a = \frac{2-\beta}{p}\K_{ab}\K^{ab}+\frac{\beta}{p}\Omega_{ab}\Omega^{ab} + \frac{\alpha}{p}(D^2 \ln p -(D\ln p)^2)-\frac{\beta}{p}(D_\perp \ln p)^2 + O(\p^3).
\]
Since the two terms $\Omega_{ab}\Omega^{ab} \geq 0$ and $-(D_\perp \ln p)^2 \leq 0$ of opposite sign might dominate the divergence when the shear tensor is small, we need $\beta =0$. Also, since the term $D^2 \ln p -(D\ln p)^2$ of indefinite sign might dominate the divergence when the shear tensor is small, we need $\alpha =0$. The entropy current $J^a = s_{eq} u^a $ is therefore the only possible expression at first order.

From the classification of second order scalars and vectors in the appendix, the general entropy current at second order takes the form
\begin{align}
\J^a &= s_{eq} u^a \Big( 1 +\frac{1}{p^2}(a_1 \K_{ab}\K^{ab}+a_2 \Omega_{ab}\Omega^{ab}+a_3 (D\ln p )^2 +a_4 D^2 \ln p +a_5 (D_\perp \ln p)^2 )\Big) \nn\\
& +\frac{s_{eq}}{p^2}\Big( b_1 h^{ac}\p_b \K^b_c +b_2 D_\perp^a D \ln p +b_3 \K^a_b D_\perp^b \ln p +b_4 \Omega^{ab}D_b^\perp \ln p +b_5 D_\perp^a \ln p D\ln p \Big),\label{J2}
\end{align}
where the coefficients $a_i$, $b_i$ are constrained by the fact that the divergence of the current is non-negative. This divergence is given by
\begin{align}
\label{bigdivJ}
\p_a \J^a &= \frac{8\pi}{p}\K_{ab}\K^{ab} +\frac{4\pi}{p^2}\Big( \K^{ab}D_a^\perp D_b^\perp \ln p (-2a_1-b_1+4b_2+b_3)+ \K^a_b \K^b_c \K^c_a(-2-2a_1+2b_2)\nn\\
& +\K^a_b \Omega^b_{\;\, c}\Omega^c_{\; \, a}(-2a_1 +4a_2 -4b_1+6 b_2)  +D\ln p \Omega_{ab}\Omega^{ab}(-4a_2+3b_1-b_2+b_4+b_5) \nn \\
&+ \p_a\K^a_c D_\perp^c \ln p (-2b_1+2b_2 +b_3+b_4)+ (D\ln p)^3(-2a_3) \nn \\
&+D\ln p \K_{ab}\K^{ab}(-2-2a_1-b_2 -b_5)+ D^3 \ln p (a_4) + D^2 \ln p D\ln p(2a_3-2a_4) \nn \\
& +\K^{ab}D^\perp_a \ln p D^\perp_b \ln p (2a_1+b_1 -3b_2 -2b_3+b_4+b_5)\nn \\
&+(D_\perp \ln p)^2 D\ln p (-2a_5-b_2-b_5)+D D^a_\perp\ln p D_a^\perp \ln p(2a_5+b_2+b_5) \Big)+O(\p^4).
\end{align}
The first term is leading in gradients and therefore the current has non-negative divergence in most cases.
It might turn out, however, that the shear tensor is small and then third order gradients might be leading.
Of these third order terms, some might form a perfect square with the term $\frac{8\pi}{p}\K_{ab}\K^{ab}$ when higher order gradients are taken into account, leading to a non-negative contribution to the divergence.  There are other terms, however, which clearly cannot form a perfect square and which may be large even though the shear is small. These latter terms must therefore vanish giving rise to the necessary positivity conditions
\bea
&& 0=-4a_2+3b_1-b_2+b_4+b_5, \label{condS1a}\\
&& 0= -2b_1+2b_2 +b_3+b_4,\label{condS1b}\\
&& 0=a_3,\label{condS1c} \\
&& 0=a_4, \label{condS1d}\\
&& 0=2a_5+b_2+b_5. \label{condS1e}
\eea
Further conditions might be found by studying the constraints up to third order or by deriving the constraints associated with putting the fluid in a curved spacetime. We will not perform such an analysis here. The constraints \eqref{condS1a}-\eqref{condS1e} should be obeyed by any physical entropy current.  Using these constraints to eliminate $a_3$, $a_4$, $a_5$, $b_4$ and $b_5$ in \eqref{J2} leads immediately  to the five-parameter family of entropy currents  with non-negative divergence \eqref{entr_curr} given in the introduction.

As a final remark, we note that certain entropy currents may be written as the divergence of an anti-symmetric potential, i.e.,
\[
\J^a = \p_b \mathcal X^{[ab]}\, .
\]
Entropy currents of this form are trivially conserved and describe lower-dimensional conservation laws at the boundary of the fluid domain. Since we consider only an infinitely extended domain, we will ignore such boundary terms. The trivial entropy currents may then be straightforwardly classified as
\[
\J^a = \p_b \left( t_1 \frac{s_{eq}}{p^2}\Omega^{ab} + t_2 \frac{2s_{eq}}{p^2}u^{[a}D_\perp^{b]}\ln p \right) \, .
\]
Making use of the relations in the appendix, one finds that the trivial entropy currents take the general form \eqref{J2} with
\bea
&&\qquad a_1 = -t_2,\qquad a_2 = -t_1+t_2,\qquad a_3 = a_4 = 0,\qquad a_5 = -t_2 , \nn\\
&& b_1 = -t_1,\qquad b_2 = -t_2,\qquad b_3 = -t_1+2t_2,\qquad b_4 = -t_1,\qquad b_5 = 3t_2\, ,
\eea
which obviously satisfy the constraints \eqref{condS1a}-\eqref{condS1e}. Removing these trivial entropy currents, one can reduce the five-parameter family of entropy currents \eqref{entr_curr} to a three-parameter family of non-trivial entropy currents.

\subsection{Defining the holographic entropy current}
\label{subsec:entropydefn}

Our bulk metric takes the form
\[
\d s^2 = -2p(x)u_a(x)\d x^a\d r + g_{ab}(r,x)\d x^a \d x^b,
\]
implying wordlines of constant $x^a$ are null geodesics to all orders. These null geodesics define a natural map between points on the horizon and on the boundary.
In the spirit of \cite{bbbb}, we will obtain a boundary entropy current by pulling back a suitable horizon quantity
along these null geodesics.
In this section we present our prescription in detail and derive a formula for the entropy current which extends that of
\cite{Booth:2010kr,Booth:2011qy} to allow the use of a non-affinely parametrised horizon generator.

We begin by foliating the bulk spacetime with a family of null hypersurfaces defined by $S(r,x)=constant$, with the
horizon corresponding to the surface $S(r_{\H}(x), x) = 0$.  In the equilibrium case, the horizon is defined to be the Rindler horizon of the bulk solution.  In the near-equilibrium case, we will assume that the fluid is returned to equilibrium in the limit of late times through the action of dissipative forces.   The horizon of the near-equilibrium solution is then defined as the unique null hypersurface which asymptotes to the Rindler horizon of the late-time equilibrium solution.
Expanding the location of the horizon $r_\H(x)$ in fluid gradients, the zeroth order term must therefore match the location of the Rindler horizon of the equilibrium solution, giving $r_\H(x) = 1-1/p^{2}+O(\p)$.  The higher order corrections may then be obtained by requiring the horizon to be null; we will return to solve for these in the next subsection.

On the horizon then, we have
\[
0 = \frac{\d S}{\d x^a}\big|_\H = \big[\p_r S \p_a r_\H + \p_a S\big]_\H.
\]
An affinely parametrised normal vector to our family of null hypersurfaces is $\ell_\mu = \p_\mu S$, since
\[
\ell^\nu\nabla_\nu\ell_\mu = \frac{1}{2}\p_\mu(\ell^2) = 0.
\]
The vanishing of $\ell^2=\ell^\mu\p_\mu S$ everywhere also implies
\[
\ell^r = -(\p_r S)^{-1}\ell^a \p_a S.
\]
Denoting the expansion evaluated on the horizon by $\theta_\H \equiv (\nabla_\mu l^\mu)_\H$, we  have
\begin{align}
\sqrt{-g}_\H\theta_\H &= \p_\mu(\sqrt{-g}\ell^\mu)|_\H = \p_a(\sqrt{-g}\ell^a)|_\H - \p_r(\sqrt{-g}(\p_r S)^{-1} \ell^a\p_a S )|_\H \nn\\[2ex]
& = \p_a[(\sqrt{-g}\ell^a)_\H]-\p_r(\sqrt{-g}\ell^a)|_\H [\p_a r_\H+(\p_r S)^{-1}\p_a S]_\H  
- \sqrt{-g_\H}\ell^a_\H\p_r((\p_r S)^{-1}\p_a S)|_\H \nn\\[2ex]
& =  \p_a[(\sqrt{-g}\ell^a)_\H]- \sqrt{-g}_\H\ell^a_\H\p_r((\p_r S)^{-1}\p_a S)|_\H,
\label{chainrule}
\end{align}
where to obtain the second line we used the chain rule in the form $\p_a[(\sqrt{-g}\ell^a)_\H] = \p_a(\sqrt{-g}\ell^a)|_\H+\p_r(\sqrt{-g}\ell^a)_\H \p_a r_\H $.

Close to the horizon, we may Taylor expand $S(r,x)$ as
\[
S(r,x) = (r-r_\H(x))S_1(x)+\frac{1}{2}(r-r_\H(x))^2 S_2(x) + O(r-r_\H)^3,
\]
where the absence of a zeroth order term is required by $S(r_\H,x)=0$.
A simple calculation shows that $S_1$ is nonzero for the Rindler horizon of the equilibrium solution;
since corrections at higher order in gradients cannot cancel this leading term, $S_1$ is everywhere nonzero in the near-equilibrium case as well.
In terms of this Taylor expansion, we obtain the {\it exact} relations
\[
\label{exact_relns}
\p_r((\p_r S)^{-1}\p_a S)|_\H = \p_a \ln S_1, \qquad
\ell^\mu_\H = S_1 (g^{\mu r}-g^{\mu b}\p_b r_\H)_\H = S_1 \xi^\mu_\H,
\]
where
\[
\label{xi_def}
\xi^\mu = g^{\mu\nu}\p_\nu(r-r_\H(x))
\]
and we have chosen $\ell^\mu_\H$ to be future-directed ensuring that $S_1>0$.  (Note $\xi^\mu_\H$ is future-directed for the equilibrium solution, and hence for the near-equilibrium solution also.)
Since $\ell^2_\H=0$, we must have $\xi^2_\H=0$. Solving this latter condition in the hydrodynamic gradient expansion provides us with the location of the horizon $r_\H(x)$, to which we will return shortly.

Combining \eqref{chainrule} and \eqref{exact_relns}, we have
\[
\sqrt{-g}_\H\theta_\H = \p_a(\sqrt{-g}_\H\ell^a_\H) - \sqrt{-g}_\H\ell^a_\H\p_a \ln S_1
= S_1\p_a \(\frac{\sqrt{-g}_\H \ell^a_\H}{S_1}\),
\]
and hence, introducing the boundary metric $g_\Sigma$, we may write
\[
\label{div_formula}
\frac{\sqrt{-g}_\H \theta_\H }{ 4G_N \sqrt{-g}_\Sigma S_1 } = \frac{1}{\sqrt{-g}_\Sigma}\,\p_a(\sqrt{-g}_\Sigma \J^a) = \nabla^{(\Sigma)}_a\J^a,
\]
where the entropy current
\[
\J^a = \frac{1}{ 4G_N} \frac{\sqrt{-g}_\H }{ \sqrt{-g}_\Sigma }\frac{\ell^a_\H }{S_1}=  \frac{1}{ 4G_N} \frac{\sqrt{-g}_\H }{ \sqrt{-g}_\Sigma }\xi^a_\H \label{ec1}\, .
\]

From the Raychaudhuri equation,
\[
\dot{\theta}_\H = -\frac{1}{d}\theta^2_\H -\sigma_{ab}\sigma^{ab}|_\H \le 0,
\]
where the dot denotes the derivative with respect to the affine parameter along the horizon and $\sigma_{ab}$ is the shear of the geodesic congruence, while the vorticity vanishes since $\ell^\mu$ is hypersurface orthogonal.
Since the fluid returns to equilibrium in the limit of late times, and $\theta_\H=0$ for the equilibrium solution,
it therefore follows that in the near-equilibrium case $\theta_\H \ge 0$ at all times.
(Note this conclusion relies on the fact that $\theta_\H$ is the expansion with respect to the affine generator $\ell^\mu_\H$.  If instead one tried to use the expansion defined with respect to the non-affine generator $\xi^\mu_\H$ then the Raychaudhuri equation would acquire extra terms of indefinite sign, invalidating the argument.)

Given then that $\theta_\H$ is non-negative and $S_1>0$, from \eqref{div_formula} the divergence of the entropy current must also be non-negative:
\[
\nabla^{(\Sigma)}_a\J^a \ge 0.
\]
Examining \eqref{ec1}, we note that while the expansion $\theta_\H$ is necessarily that of the affinely parametrised generator $\ell^\mu_\H$, the current $\J^a$ may nevertheless be expressed in terms of the non-affinely parametrised generator $\xi_\H^\mu$.

Finally, let us discuss briefly two potential sources of ambiguity in the definition of the holographic entropy current.  Firstly, pulling back to the boundary along a different set of bulk null geodesics will lead to a different boundary entropy current. Such ambiguities have been discussed in \cite{bbbb} and correspond
boundary to boundary diffeomorphisms. A second potential source of ambiguity, mentioned in \cite{bbbb} and discussed in
\cite{Booth:2010kr,Booth:2011qy}, concerns the choice of bulk
horizon (such as apparent horizon, etc.). We leave further investigation of these interesting issues to future
work, and in the following, we focus exclusively on the entropy current defined in \eqref{ec1}.

\subsection{Location of the horizon}
\label{sec:posh}

To evaluate the entropy current according to our formula \eqref{ec1}, the first step is to compute $r_\H(x)$, the location of the horizon.
As explained above, this follows from solving the null condition $\xi^2_\H=0$, which from \eqref{xi_def} reads
\[
0=g^{rr}(r_\H) - 2 g^{ra}(r_\H)\p_a r_\H + g^{ab}(r_\H) \p_a r_\H \p_b r_\H.\label{defh2}
\]
The solution takes the form of a gradient expansion
\[
r_\H(x) = r_\H^{(0)} + r_\H^{(1)}+r_\H^{(2)}+\dots \label{def:rh}
\]
where $r_\H^{(0)} = 1-1/p^{2}$ is the equilibrium position of the horizon obtained by solving
\[
g_{(0)}^{rr}(r^{(0)}_\H) = p^{-2}+r^{(0)}_\H-1=0\, ,
\]
and $r_\H^{(n)}$ contains terms of $n$-th order in gradients. At each order $n$ in gradients, the equation reduces to a linear problem due to the fact that the only term involving $r_\H^{(n)}$ is $g_{(0)}^{rr}(r^{(n)}_\H) = r_\H^{(n)}$.

At first order, \eqref{defh2} reads
\[
0=r_\H^{(1)}  - \frac{1}{p^2}\alpha^{(1)}(r_\H^{(0)})-\frac{2}{p} u^a\p_a r_\H^{(0)}\, ,
\]
with solution
\bea
r_\H^{(1)} = \frac{2}{p^3}D\ln p\, .
\eea
At second order, we obtain the equation
\begin{align}
0 & =  r_\H^{(2)} - \frac{2}{p}D\ln p \, r_\H^{(1)} - \frac{1}{p^2}\alpha^{(2)}(r_\H^{(0)}) + \frac{4}{p^2}(D_\perp \ln p)^2 -\frac{2}{p}D r_\H^{(1)} \nn \\&
\quad -\frac{4}{p^2} D_\perp^a \ln p D^\perp_a r_\H^{(0)} + h^{ab}\p_a r_\H^{(0)}\p_b r_\H^{(0)} \, ,
\end{align}
with solution
\bea
r_\H^{(2)} = \frac{1}{p^4}\Big( 4 DD \ln p -8 (D\ln p )^2 - \frac{3}{2}\mathcal K_{ab}\mathcal K^{ab}-\frac{1}{2}\Omega_{ab}\Omega^{ab}+(D_\perp \ln p )^2 \Big)\, .
\label{valrh2}
\eea

Having obtained $r_\H(x)$, we may now evaluate the non-affine horizon generator $\xi^\mu_\H$ according to \eqref{xi_def}, making use of the inverse metric \eqref{invg2} evaluated on the horion.  We find
\[
\xi_\H^a =\frac{u^a}{p}+\frac{1}{p^3} \Big( 2 D_\perp^a \ln p D\ln p -2 D^a_\perp D\ln p - h^{ab} \p_c \mathcal K^c_b + (\mathcal K^a_{\;\, b}+\Omega^a_{\; \, b})D_\perp^b \ln p \Big)+ O(\p^3).\label{valxi2}
\]
We will not need $\xi^r_\H$ in what follows. The piece of $\xi_\H^a$ normal to the fluid velocity is simply the equilibrium term, with no corrections at either first or second order.  The remaining piece of $\xi_\H^a$ tangent to the fluid velocity has only second order corrections.

\subsection{Evaluating the holographic entropy current}

In our case the boundary metric is simply Minkowski and $1/4G_N = 4\pi$, so the entropy current \eqref{ec1} reduces to
\[
\J^a =4\pi \sqrt{-g}_\H \xi^a_\H.
\]
With $\xi_\H^a$ given in \eqref{valxi2} above, it remains only to evaluate the determinant factor $\sqrt{-g}_\H$.

As an initial step, we first evaluate the determinant of the seed metric
\[
\d s_{(0)}^2 = -2pu_a\d x^a \d r+ \bg_{ab}\d x^a \d x^b,
\]
where, in the above and in the following, for clarity we will temporarily write
\[
\bg_{ab} = g^{(0)}_{ab} = \eta_{ab}-p^2(r-1)u_au_b\,.
\]
Since
\[
\(\begin{array}{cc} 0 & -p{\bf u}^T \\ -p {\bf u} & {\bf \bg} \end{array}\)
=\(\begin{array}{cc} 1 & -p{\bf u}^T \\ {\bf 0} & {\bf \bg} \end{array}\)
\(\begin{array}{cc} -p^2 {\bf u}^T{\bf \bg^{-1} u} & {\bf 0} \\ -p {\bf \bg^{-1} u} & \text{{\bf 1}} \end{array}\),
\]
we find
\[
\det g_{(0)} = -p^2 u_a u_b \bg^{ab} \det \bg,
\]
where the inverse metric
\[
\bg^{ab} = h^{ab}-\frac{u^a u^b}{(1+p^2(r-1))}, \qquad
\bg^{ab}\bg_{bc} = \delta^a_c,
\]
and the determinant $\det \bg = -(1+p^2(r-1))$ may be evaluated by the usual formula. The seed metric therefore has determinant
\[
\label{seeddet}
\det g_{(0)} = -p^2.
\]

The determinant of the full metric up to second order in gradients may now be obtained perturbatively.
Writing
\[
g_{\mu\nu} = g_{(0)\mu\nu}+g_{(1)\mu\nu}+g_{(2)\mu\nu} + O(\p^3)
\]
and expanding the formula $\det g = \exp(\mathrm{tr}\ln g)$,
we find
\[
\det g = \det g_{(0)} \(1+ g_{(0)}^{\mu\nu}g_{(1)\mu\nu} +  g_{(0)}^{\mu\nu}g_{(2)\mu\nu} - \frac{1}{2}g_{(0)}^{\mu\nu}g_{(1)\nu\rho}g^{\rho\sigma}_{(0)}g_{(1)\sigma\mu} + \frac{1}{2}(g^{\mu\nu}_{(0)}g_{(1)\mu\nu})^2\) + O(\p^3).
\]
Since in addition,
\[
g_{(1)r\mu}=g_{(2)r\mu}=0, \quad  g^{ab}_{(0)}=h^{ab},\quad \gamma_{(1)ab}\equiv h_a^c h_b^d g_{(1)cd} =0,
\]
making use of \eqref{seeddet}, we find simply
\[
\det g = -p^2 (1+ h^{ab}\g_{(2)ab}) + O(\p^3).
\]
Evaluating this formula on the horizon $r_\H = 1-1/p^2 + O(\p)$, we obtain
\[
\sqrt{-g}_\H = p + \frac{1}{p}\cK_{ab}\cK^{ab} + \frac{1}{2p}\Omega_{ab}\Omega^{ab} + O(\p^3),
\]
and thus the entropy current
\begin{align}
\J^a &=
4\pi u^a\(1+\frac{1}{p^2}\cK_{bc}\cK^{bc}+\frac{1}{2p^2}\Omega_{bc}\Omega^{bc}\) \nn\\&\quad
+\frac{4\pi}{p^2}\(2D^a_\perp \ln p D \ln p - 2 D^a_\perp D\ln p - h^{ab}\p_c \cK^c_b+(\cK^a_b+\Omega^a_{\,\,b})D^b_\perp\ln p\) + O(\p^3).
\end{align}
The entropy current takes the general form \eqref{J2} with coefficients
\bea
a_1 = 1,\quad a_2 = \frac{1}{2},\quad a_3=a_4=a_5 = 0,\quad -b_1 = b_3=b_4 = 1,\quad -b_2 = b_5 = 2.\label{valai}
\eea

From \eqref{bigdivJ}, we obtain
\begin{align}
\p_a \J^a = \frac{8\pi}{p}\K_{ab}\K^{ab} +\frac{4\pi}{p^2}\Big(& -8 \K^{ab}D_a^\perp D_b^\perp \ln p -8 \K^a_b \K^b_c \K^c_a -8 \K^a_b \Omega^b_{\;\, c}\Omega^c_{\; \, a} \nn\\&
+ 8\K^{ab}D^\perp_a \ln p D^\perp_b \ln p -4  D\ln p \K_{ab}\K^{ab}\Big) + O(\p^4).
\end{align}
All five entropy conditions \eqref{condS1a}-\eqref{condS1e} are obeyed, confirming that the divergence of the entropy current is non-negative as expected.

\section{Near-horizon limits} \label{sec:near}

It was observed in \cite{Bredberg} that the non-relativistic hydrodynamic expansion can be expressed as a near-horizon limit when combined with a specific Weyl rescaling. In this section we show that the relativistic expansion can also be expressed as an alternative near-horizon limit combined with a Weyl rescaling.

We noticed in section \ref{sec:seed} that the equilibrium solution admits the scaling transformation \eqref{rescale} that generates a global Weyl rescaling of the metric. This Weyl rescaling still exists for the complete metric with higher-derivative corrections that we found in section \ref{sec:sol} (with the factors of $r_c$ restored). Indeed, the metric has coordinates $r,\t,x_i$ and parameters $r_c$, $p(\tau,x)$ and $v_i(\t,x)$, where the relativistic velocity $u^a$ is decomposed as in \eqref{relueq}. Equivalently, one can express the solution using the position of the horizon $r_\H(\t,x)$ instead of the pressure after inverting the relation \eqref{def:rh}. The scaling
\[
\label{Weyl}
(r,\t,x_i,r_c,r_\H,v_i) \tto \( \lambda^2 \, r , \t, \lambda\, x_i , \lambda^2 \, r_c ,\lambda^2 \, r_\H , \lambda\, v_i \)
\]
is equivalent to a Weyl rescaling
\[
\d s^2 \tto \lambda^2\, \d s^2
\]
of the full near-equilibrium metric. Since we are interested in Ricci-flat metrics this constant overall factor may be dropped.

We now want to consider the near-horizon limit $r_c \to r_\H \to 0$ while preserving $r_\H<r_c$ and $r_c-v^2>0$ (recall that
$\sqrt{r_c}$ also plays the role of the speed of light). Thus, in general we must scale the parameters
\[
\label{NH}
(r_c, r_\H, v_i) \tto (\tilde{\lambda} r_c, \tilde{\lambda}^a r_\H, \tilde{\lambda}^b v_i)
\]
such that $a\ge 1$ and $b\ge 1/2$.  According to this point of view, any suitable choice of $a$ and $b$ defines an acceptable near-horizon limit.

Recall that the non-relativistic $\ep$-expansion -- defined as the homogenous scaling transformation of the incompressible Navier-Stokes equations -- is given by
\[
\label{ep}
(r,\t,x_i,r_c,r_\H,v_i) \tto \(r, \frac{\t}{\ep^2},\frac{x_i}{\ep},r_c,\ep^2 r_\H, \ep v_i\)
\]
Recall also that the incompressible Navier-Stokes equations are an attractor under the $\ep$ scaling in the sense that when $\ep \rightarrow 0$ all higher order corrections become small.

Combining this expansion with a Weyl rescaling (\ref{Weyl}), setting $\lambda=\ep$, leads to
\[
\label{NHnonrel}
(r,\t,x_i,r_c,r_\H,v_i) \tto \(\ep^2 r, \frac{\t}{\ep^2},x_i,\ep^2r_c, \ep^4 r_\H, \ep^2v_i\),
\]
which defines a near-horizon limit (\ref{NH}) with $\tilde{\lambda}=\ep^2$ and $a=2$, $b=1$, i.e., we consider the limit
\[ \label{NRlimit}
r_c\tto 0, \qquad  \frac{r_\H}{r_c^2} = {\rm fixed}, \qquad  \frac{v_i}{r_c}={\rm fixed}.
\]

Let us now consider the relativistic limit
\[
\label{rel}
(r,\t,x_i,r_c,r_\H,v_i) \tto \(r,\frac{\t}{\tilde{\ep}},\frac{x_i}{\tilde{\ep}},r_c,r_\H,v_i\),
\]
i.e., the pressure $p$ and $u^a$ are zeroth order quantities, only derivatives carry weight $\p_\mu \sim \tilde{\ep}$. The ideal relativistic fluid equations \eqref{eq:zeroth} are an attractor under the relativistic scaling in the sense that when $\tilde\ep \rightarrow 0$, all higher order corrections become small.

Combining this expansion with a Weyl rescaling (\ref{Weyl}) with $\lambda=\tilde{\ep}$ leads to the near-horizon limit
\[
\label{NHrel}
(r,\t,x_i,r_c,r_\H,v_i) \tto \(\tilde{\ep}^2 r, \frac{\t}{\tilde{\ep}},x_i,\tilde{\ep}^2 r_c, \tilde{\ep}^2r_\H, \tilde{\ep} v_i\),
\]
which is (\ref{NH}) with $\tilde{\lambda}=\tilde{\ep}^2$ and $a=1$ and $b=1/2$, i.e., we consider the limit
\[ \label{Rlimit}
r_c\tto 0, \qquad  \frac{r_\H}{r_c} = {\rm fixed}, \qquad  \frac{v^2}{r_c}={\rm fixed}.
\]
In particular, this means that we keep fixed relativistic velocities as $\sqrt{r_c}$ plays the role of the speed of light. Notice that  under (\ref{NHrel}) the normalised horizon generator $\zeta = \frac{1}{\sqrt{r_c-v^2}}(\p_\t+v_i\p_i)$ (where $\zeta^2|_{\Sigma_c} = -1$, see \cite{firstpaper}), is invariant and the temperature and pressure
satisfy a simple scaling law $T \to T/\tilde{\ep}$ and $p \to p/\tilde{\ep}$, while the transformations under the non-relativistic scaling (\ref{NHnonrel}) are
more complicated.

\section{Conclusions} \label{sec:disc}

In this paper we presented a construction of a $(d+2)$-dimensional Ricci-flat metric corresponding to a $(d+1)$-dimensional relativistic fluid with specific transport coefficients. In a specific non-relativistic limit we recover the results discussed in our previous work \cite{firstpaper}. We have further obtained a holographic entropy current with a non-negative divergence, in accordance with the second law of thermodynamics. We also showed how to reinterpret the relativistic hydrodynamic expansion as certain near-horizon limit.

There are many interesting directions that one may wish to pursue further. Some of the numerous questions that were raised in \cite{firstpaper} have now been addressed, both in the literature discussed in the introduction, and in the present work.  Nonetheless, many interesting questions remain. Perhaps most far-reaching of these are the questions concerning holography. How concrete can we make this holographic duality?  Can we move away from the hydrodynamic regime? Can we set up holography for general spacetimes by using the discussion here as a local holographic reconstruction of small neighbourhoods which should then be patched together to obtain a global description? Answering any of these questions would be a significant step towards formulating a general theory of holography.

\section*{Acknowledgments}

We thank Michal Heller and Christopher Eling for discussions.
This work is part of the research program of the Stichting voor
Fundamenteel Onderzoek der Materie (FOM), which is financially
supported by the  Nederlandse Organisatie voor Wetenschappelijk
Onderzoek (NWO). KS and GC acknowledge support via an NWO
Vici grant, and PM via an NWO Veni grant.
Research at the Perimeter Institute is supported by the Government of Canada
through Industry Canada and by the Province of Ontario through the Ministry of
Research \& Innovation.

\appendix

\section{Basis of hydrodynamic scalars and vectors}
\label{app:basis}

At equilibrium, the long wavelength dynamics of relativistic quantum field theories in flat spacetime at vanishing
charge density can usually be described by a set of fundamental fluid variables
consisting of the energy density $\rho$, pressure $p$ and the fluid vector $u^a$, constrained by an equation of state $\rho = \rho(p)$.
For many applications it is usually more convenient and natural
to trade the pressure for the temperature $T$ and treat $(T,u^a)$ as the fundamental variables.
At non-zero energy density either choice of fundamental variables is equally valid but the equation of state of the equilibrium fluid dual
to vacuum Einstein gravity is $\rho = 0$. Conservation of the stress tensor $T_{ab} = p h_{ab}+O(\p)$ then leads to the incompressible ideal fluid equations
\bea
\p_a u^a = 0+O(\p^2) ,\qquad a^a = - D^\perp_a \ln p +O(\p^2)\, .\label{eq:A1}
\eea
The fluid is incompressible at first order in gradients precisely because the equilibrium energy density is zero.
As a consequence, the bases of hydrodynamic scalars and vectors traditionally used to describe the fluid dynamics
at higher orders in gradients, see e.g.~\cite{Romatschke}, are not applicable to this special case.
In this appendix we construct a convenient basis for the hydrodynamics of fluids with zero equilibrium energy density, using
$(p,u^a)$ as the fundamental variables, and
we provide the relations which can be used to express other linearly dependent fluid scalars (up to third order in gradients),
vectors and tensors (up to second order in gradients) in terms of this basis.

At zeroth order, there is only one scalar, $p$, and one vector, $u^a$, along with
one symmetric tensor orthogonal to $u^a$, $h_{ab} \equiv \eta_{ab}+u_a u_b$.
At first order in gradients, the scalar $\p_a u^a$, or equivalently $\K^a_a$, is higher order in gradients
due to the incompressibility equation. Therefore $D \ln p$ is the only independent scalar at this order. The vectors orthogonal to $u^a$
are the acceleration $a^a$ and the pressure gradient orthogonal to the fluid velocity $D^a_\perp \ln p$. However, the equations of motion
imply that only one of them, which we choose to be $D^a_\perp \ln p$, is independent.
The symmetric tensors orthogonal to $u^a$ are $\K_{ab}$ and $D\ln p\,  h_{ab}$. The latter is isotropic, i.e., proportional to $h_{ab}$.
There is an one-to-one mapping between non-isotropic symmetric tensors and traceless symmetric tensors. Isotropic symmetric tensors
(proportional to $h_{ab}$) may be classified by their multiplicative prefactor. Putting together these considerations, one can derive a
basis for scalars, vectors and tensors up to first order in gradients, which is summarised in Table \ref{table1}.
Dependent scalars and vectors can be expressed in terms of this basis using the first order equations of motion given in
\eqref{eq:A1}. Note that derivatives of the fluid velocity $u^a$ can be expressed as
\bea
\p_a u_b  = \K_{ab}+\Omega_{ab} - u_a a_b\, .
\eea

\begin{table}[t]\begin{center}
\begin{tabular}{|l|l|l|}\hline
Order in gradients & 0 & 1 \\ \hline
Scalars & $p$ & $D\ln p $ \\ \hline
Vectors orthogonal to $u^a$ & -- & $D^a_\perp \ln p $ \\ \hline
Symmetric non-isotropic tensors orthogonal to $u^a$ & -- & $\K_{ab}$ \\ \hline
\end{tabular}\end{center}\vspace{-2ex}\caption{Basis of scalars, vectors and tensors at zeroth and first order in gradients.}\label{table1}
\end{table}
At second order in derivatives, one can use the fluid equations \eqref{eq:A1} and their first derivative to derive the following relationships:
\begin{align}
\p_a D_\perp^a \ln p  &= -\K_{ab} \K^{ab} + \Omega_{ab}\Omega^{ab}+O(\p^3),\label{rel1} \\
D^\perp_a D_\perp^a \ln p  &= h^{ab}\p_a \p_b \ln p +O(\p^3) = -\K_{ab} \K^{ab} + \Omega_{ab}\Omega^{ab}+(D_\perp \ln p )^2 +O(\p^3), \label{rel2}\\
 u^c u^d \p_c \p_d \ln p &= D D \ln p + (D_\perp \ln p)^2  + O(\p^3), \label{rel3} \\
 h^ c_a \p_b \Omega^b_{\; \,c} &= h^c_a \p_b \mathcal K^b_{\; c}+(\mathcal K_{ab}-\Omega_{ab})D_\perp^b \ln p +O(\p^3), \\
 h^{cd}D\mathcal K_{cd} &= D\mathcal K = O(\p^3), \\
 h_{(a}^c D_{b)}^\perp D^\perp_c \ln p &=  h_{a}^c h_{b}^d \p_c \p_d \ln p +  \mathcal K_{ab}D \ln p +O(\p^3),  \\
 \p_{(a}D_{b)}^\perp \ln p &= u_a u_b (D_\perp \ln p )^2 + \K_{ab} D\ln p +h_a^c h_b^d \p_c\p_d \ln p \nn\\
&\quad + u_{(a}\Big( 2(K_{b)c}+\Omega_{b)c})D^c_\perp \ln p - D^\perp_{b)}D\ln p +D^\perp_{b)}\ln p D \ln p\Big) +O(\p^3), \\
h^c_a h^d_b D\mathcal K_{cd} &= - h_{a}^c h_{b}^d \p_c \p_d \ln p - \mathcal K_{ab}  D \ln p + D^\perp_a \ln p D^\perp_b \ln p 
-\mathcal K_a^{\; c}\mathcal K_{cb}  -\Omega_a^{\; \, c}\Omega_{cb} \nonumber \\& \quad
+O(\p^3).\label{rel4}
\end{align}
Note also the following exact relation
\bea
\mbox{}[ D, D^\perp_a ]  &=& a_a D + u_a a^b D^\perp_b - (\mathcal K_a^b +\Omega_a^{\;\, b})D^\perp_b \label{rel4b}\, .
\eea
Taking these relations into account, we can choose as a basis the five scalars and the five vectors
orthogonal to the fluid velocity $u^a$ that are indicated in Table~\ref{table2}. One also finds six independent symmetric non-isotropic tensor fields orthogonal to $u^a$. Dependent quantities may be expressed in terms of this basis using the relationships given above.
Our result is consistent with that of \cite{Romatschke}:
the number of independent fields at second order coincides (even though the basis of scalars, vectors and tensors is different).
\begin{table}[t]\begin{center}
\begin{tabular}{|l|c|}\hline
Order in gradients & 2 \\ \hline
Scalars & $\K_{ab}\K^{ab}$, $\Omega_{ab}\Omega^{ab}$, $(D \ln p )^2$, $D D \ln p$, $(D_\perp \ln p )^2$ \\ \hline
Vectors orthogonal to $u^a$ & $h_a^c \p_b \K^b_c$, $D^\perp_a D \ln p $, $\K_{ab}D^b_\perp \ln p $, $\Omega_{ab} D^b_\perp \ln p $, $D^\perp_a\ln p D\ln p $\\ \hline
Symmetric non-isotropic tensors  &
$\K_a^c\K_{cb} $, $\K_{(a}^c\Omega_{|c|b)}$, $\Omega_a^{\,\,\,c}\Omega_{cb}$, $h_a^ch_b^d\D_c\D_d \ln p$, $\K_{ab}\,D\ln p$, \\
orthogonal to $u^a$ &  $D^\perp_a \ln p \,D^\perp_b\ln p $\\ \hline
\end{tabular}\end{center}\vspace{-2ex}\caption{Basis of scalars, vectors and tensors at second order in gradients.}\vspace{2ex}\label{table2}
\end{table}

At third order in gradients, we obtain the following relationships using the fluid equations \eqref{eq:A1} and their first and second derivatives:
{\allowdisplaybreaks
\begin{align}
\K^{ab}D \K_{ab} &= -\K^{ab}D^\perp_a D^\perp_b \ln p+ \K^{ab}D^\perp_a \ln p D^\perp_b \ln p - \K^a_b \K^b_c \K^c_a - \K^a_b \Omega^b_{\; \, c}\Omega^c_{\;\, a} \nonumber \\&\quad 
+O(\p^4),\label{rel5}\\
\Omega^{ab}D \Omega_{ab} &= -  \Omega_{ab}\Omega^{ab} D \ln p+2 \K^a_b \Omega^b_{\; \, c}\Omega^c_{\;\, a} +O(\p^4),\label{rel5b}\\
D \p_a D_\perp^a \ln p &= 2\K^{ab}D^\perp_a D^\perp_b \ln p-2 \K^{ab}D^\perp_a \ln p D^\perp_b \ln p +2 \K^a_b \K^b_c \K^c_a +6 \K^a_b \Omega^b_{\; \, c}\Omega^c_{\;\, a}\nn\\
&\quad -2\Omega_{ab}\Omega^{ab} D\ln p +O(\p^4) ,\label{rel6}\\
\mathcal K^{ab} \p_a \p_b \ln p &= \K^{ab} D^\perp_a D^\perp_b \ln p - \K_{ab}\K^{ab} D\ln p +O(\p^4) ,\\
\p_a \Omega^a_{\; \, b}D_\perp^b \ln p &= \p_a \K^a_{\; \, b}D_\perp^b \ln p + \K^{ab}D_a^\perp \ln p D_b^\perp \ln p +O(\p^4) ,\\
\Omega^{ab}D^\perp_a D^\perp_b \ln p &= \Omega_{ab}\Omega^{ab} D\ln p +O(\p^4) ,\\
D_\perp^a \ln p D^\perp_a D \ln p &= D_\perp^a \ln p D D^\perp_a \ln p + (D_\perp \ln p )^2 D\ln p  
+ \K^{ab}D^\perp_a \ln p D^\perp_b \ln p \nonumber \\& \quad
+O(\p^4),\\ \nonumber 
\p_a D_\perp^a D \ln p &=
4 \K^{ab} D^\perp_a D^\perp_b \ln p -  \K^{ab}D_a^\perp \ln p D_b^\perp \ln p+2\K^a_b \K^b_c \K^c_a+6\K^a_b \Omega^b_{\; \, c}\Omega^c_{\;\, a} \nn \\
&\quad - \K_{ab}\K^{ab} D\ln p - \Omega_{ab}\Omega^{ab} D\ln p+3 D_\perp^a \ln p DD^\perp_a \ln p
\nn \\ &\quad  +(D_\perp \ln p)^2 D\ln p 
+ 2 \p_a \K^a_{\; \, b}D_\perp^b \ln p +O(\p^4), \label{rel7}\\
D_\perp^b \p_a \K^a_{b} &= -4  \K^a_b \Omega^b_{\;\, c}\Omega^c_{\; \, a} + \p_a \K^{ab} D^\perp_b \ln p +\mathcal K^{ab}D^\perp_a \ln p D^\perp_b \ln p \nn\\
&\quad + 3\Omega_{ab}\Omega^{ab} D\ln p -\K^{ab}D_a^\perp D_b^\perp \ln p  +O(\p^4) . \label{rel8}
\end{align}
}
Note that the higher order correction terms to the fluid equations \eqref{eq:A1}
only appear at subleading orders in gradients in these relations.
\begin{table}[t]\begin{center}
\begin{tabular}{|l|c|}
\hline
Order in gradients & 3 \\ \hline
Scalars & $\K^{ab}D_a^\perp D_b^\perp \ln p$,  $ \K^a_b \K^b_c \K^c_a$, $\K^a_b \Omega^b_{\;\, c}\Omega^c_{\; \, a}$, $D\ln p \Omega_{ab}\Omega^{ab}$, $ \p_a\K^a_c D_\perp^c \ln p$,  \\
& $\K^{ab}D^\perp_a \ln p D^\perp_b \ln p$, $D\ln p \K_{ab}\K^{ab} $, $D^3 \ln p $, $D^2 \ln p D\ln p$, $(D\ln p)^3$\\
& $(D_\perp \ln p)^2 D\ln p$, $D D^a_\perp\ln p D_a^\perp \ln p$\\
\hline
\end{tabular}\end{center}\vspace{-2ex}\caption{Basis of scalars at third order in gradients.}\label{table3}
\end{table}
Taking these relations into account, we may choose a basis of twelve scalars\footnote{{\bf Note added:} These twelve scalars appear to be consistent with the classification obtained recently in \cite{Bhattacharyya:2012ex}.
Indeed, according to Tables 4, 8, 9 and 10 of \cite{Bhattacharyya:2012ex}, at third order there is one scalar involving three derivatives of a zeroth order quantity, four scalars involving the product of a one-derivative term and a two-derivative term, and seven scalars representing products of three one-derivative terms.
Note however that the results of \cite{Bhattacharyya:2012ex} require the use of Landau gauge, and so are not directly applicable to the Rindler fluid with vanishing equilibrium energy density considered here.} at third order, as indicated
in Table~\ref{table3}.


\begin{thebibliography}{99}


\bibitem{Bredberg}
  I.~Bredberg, C.~Keeler, V.~Lysov and A.~Strominger,
  ``From Navier-Stokes To Einstein'',
  [arXiv:1101.2451].

\bibitem{firstpaper}
  G.~Comp\`ere, P.~McFadden, K.~Skenderis and M.~Taylor,
  ``The Holographic fluid dual to vacuum Einstein gravity,''
  JHEP {\bf 1107} (2011) 050
  [arXiv:1103.3022 [hep-th]].


\bibitem{Bredberg:2011xw}
  I.~Bredberg and A.~Strominger,
  ``Black Holes as Incompressible Fluids on the Sphere,''
  arXiv:1106.3084 [hep-th].

\bibitem{Huang:2011he}
  T.~-Z.~Huang, Y.~Ling, W.~-J.~Pan, Y.~Tian and X.~-N.~Wu,
  ``From Petrov-Einstein to Navier-Stokes in Spatially Curved Spacetime,''
  JHEP {\bf 1110}, 079 (2011)
  [arXiv:1107.1464 [gr-qc]].

\bibitem{Nakayama:2011bu}
  R.~Nakayama,
  ``The Holographic Fluid on the Sphere Dual to the Schwarzschild Black Hole,''
  arXiv:1109.1185 [hep-th].


\bibitem{Anninos:2011zn}
  D.~Anninos, T.~Anous, I.~Bredberg and G.~S.~Ng,
  ``Incompressible Fluids of the de Sitter Horizon and Beyond,''
  arXiv:1110.3792 [hep-th].

\bibitem{Chirco:2011ex}
  G.~Chirco, C.~Eling and S.~Liberati,
  ``Higher Curvature Gravity and the Holographic fluid dual to flat spacetime,''
  JHEP {\bf 1108}, 009 (2011)
  [arXiv:1105.4482 [hep-th]].

\bibitem{Niu:2011gu}
  C.~Niu, Y.~Tian, X.~-N.~Wu and Y.~Ling,
  ``Incompressible Navier-Stokes Equation from Einstein-Maxwell and Gauss-Bonnet-Maxwell Theories,''
  arXiv:1107.1430 [hep-th].


\bibitem{Damour1}
T.~Damour,
``Quelques propri{\'e}t{\'e}s m{\'e}caniques, {\'e}lectromagn{\'e}tiques, thermodynamiques et
quantiques des trous noirs,'' Th{\`e}se de Doctorat d'Etat, Universit{\'e} Pierre et Marie Curie, Paris VI,
1979.

\bibitem{Damour2}
T.~Damour,
``Surface effects in black hole physics,''
Proceedings of the second Marcel Grossmann meeting on General Relativity (1982),
Ed.~R.~Ruffini, North Holland.

\bibitem{Thorne}
K.~S.~Thorne, R.~H.~Price, and D.~A.~Macdonald,
``Black Holes: the Membrane Paradigm,'' Yale University Press,  New
Haven, USA (1986).


\bibitem{Bhattacharyya:2008jc}
  S.~Bhattacharyya, V.~E.~Hubeny, S.~Minwalla and M.~Rangamani,
  ``Nonlinear Fluid Dynamics from Gravity,''
  JHEP {\bf 0802} (2008) 045,
  [arXiv:0712.2456].

\bibitem{Fouxon:2008tb}
  I.~Fouxon and Y.~Oz,
  ``Conformal Field Theory as Microscopic Dynamics of Incompressible Euler and
  Navier-Stokes Equations,''
  Phys.\ Rev.\ Lett.\  {\bf 101} (2008) 261602,
  [arXiv:0809.4512].

\bibitem{Bhattacharyya:2008kq}
  S.~Bhattacharyya, S.~Minwalla and S.~R.~Wadia,
  ``The Incompressible Non-Relativistic Navier-Stokes Equation from Gravity,''
  JHEP {\bf 0908} (2009) 059,
  [arXiv:0810.1545].

\bibitem{Eling:2009pb}
  C.~Eling, I.~Fouxon and Y.~Oz,
  ``The Incompressible Navier-Stokes Equations From Membrane Dynamics,''
  Phys.\ Lett.\  B {\bf 680} (2009) 496,
  [arXiv:0905.3638].


\bibitem{Policastro:2001yc}
  G.~Policastro, D.~T.~Son and A.~O.~Starinets,
  ``The Shear viscosity of strongly coupled N=4 supersymmetric Yang-Mills plasma,''
  Phys.\ Rev.\ Lett.\  {\bf 87}, 081601 (2001)
  [hep-th/0104066].


\bibitem{Emparan:2009cs}
  R.~Emparan, T.~Harmark, V.~Niarchos and N.~A.~Obers,
  ``World-Volume Effective Theory for Higher-Dimensional Black Holes,''
  Phys.\ Rev.\ Lett.\  {\bf 102}, 191301 (2009)
  [arXiv:0902.0427 [hep-th]].

\bibitem{Emparan:2011hg}
  R.~Emparan, T.~Harmark, V.~Niarchos and N.~A.~Obers,
  ``Blackfolds in Supergravity and String Theory,''
  JHEP {\bf 1108}, 154 (2011)
  [arXiv:1106.4428 [hep-th]].


\bibitem{Cai:2011xv}
  R.~-G.~Cai, L.~Li and Y.~-L.~Zhang,
  ``Non-Relativistic Fluid Dual to Asymptotically AdS Gravity at Finite Cutoff Surface,''
  JHEP {\bf 1107}, 027 (2011)
  [arXiv:1104.3281 [hep-th]].


\bibitem{Mei:2011gv}
  J.~Mei,
  ``Towards A Possible Fluid Flow Underlying the Kerr Spacetime,''
  arXiv:1104.3728 [hep-th].

\bibitem{Lysov:2011xx}
  V.~Lysov and A.~Strominger,
  ``From Petrov-Einstein to Navier-Stokes,''
  arXiv:1104.5502 [hep-th].

\bibitem{Kuperstein:2011fn}
S.~Kuperstein and A.~Mukhopadhyay,
``The unconditional RG flow of the relativistic holographic fluid,'' JHEP {\bf 1111} (2011) 130 [arXiv:1105.4530].

\bibitem{Brattan:2011my}
  D.~K.~Brattan, J.~Camps, R.~Loganayagam and M.~Rangamani,
  ``CFT dual of the AdS Dirichlet problem : Fluid/Gravity on cut-off surfaces,''
  arXiv:1106.2577 [hep-th].


\bibitem{Verschelde:2011wr}
  H.~Verschelde and V.~I.~Zakharov,
  ``Notes on relativistic superfluidity and gauge/string duality,''
  arXiv:1106.4154 [hep-th].

\bibitem{Eling:2011ct}
  C.~Eling and Y.~Oz,
  ``Holographic Screens and Transport Coefficients in the Fluid/Gravity Correspondence,''
  Phys.\ Rev.\ Lett.\  {\bf 107}, 201602 (2011)
  [arXiv:1107.2134 [hep-th]].

\bibitem{Rodrigues:2011gg}
  F.~G.~Rodrigues, W.~A.~Rodrigues, Jr and R.~da Rocha,
  ``Maxwell and Navier-Stokes Equations Equivalent to Einstein Equation,''
  arXiv:1109.5274 [math-ph].


\bibitem{Huang:2011kj}
  T.~-Z.~Huang, Y.~Ling, W.~-J.~Pan, Y.~Tian and X.~-N.~Wu,
  ``Fluid/Gravity duality with Petrov boundary condition in a spacetime with a cosmological constant,''
  arXiv:1111.1576 [hep-th].


\bibitem{Marolf:2012dr}
  D.~Marolf and M.~Rangamani,
  ``Causality and the AdS Dirichlet problem,''
  arXiv:1201.1233 [hep-th].

\bibitem{Witten1998}
  E.~Witten,
  ``Anti-de Sitter space, thermal phase transition, and confinement in  gauge
  theories,''
  Adv.\ Theor.\ Math.\ Phys.\  {\bf 2} (1998) 505
  [arXiv:hep-th/9803131].

\bibitem{bbbb}
  S.~Bhattacharyya, V.~E.~Hubeny, R.~Loganayagam, G.~Mandal, S.~Minwalla,
T.~Morita, M.~Rangamani and H.~S.~Reall,
  ``Local Fluid Dynamical Entropy from Gravity,''
  JHEP {\bf 0806}, 055 (2008)
  [arXiv:0803.2526 [hep-th]].

\bibitem{Elingtoappear}
  C. Eling, A. Meyer and Y. Oz, ``The Relativistic Rindler Hydrodynamics'', arXiv:1201.2705.

\bibitem{Papadimitriou:2004ap}
  I.~Papadimitriou and K.~Skenderis,
  ``AdS / CFT correspondence and geometry,''
  hep-th/0404176.



\bibitem{Booth:2010kr}
  I.~Booth, M.~P.~Heller and M.~Spalinski,
  ``Black Brane Entropy and Hydrodynamics,''
  Phys.\ Rev.\  D {\bf 83} (2011) 061901
  [arXiv:1010.6301 [hep-th]].


\bibitem{Booth:2011qy}
  I.~Booth, M.~P.~Heller, G.~Plewa and M.~Spalinski,
  ``On the apparent horizon in fluid-gravity duality,''
  Phys.\ Rev.\  D {\bf 83} (2011) 106005
  [arXiv:1102.2885 [hep-th]].




\bibitem{Romatschke}
  P.~Romatschke,
  ``Relativistic Viscous Fluid Dynamics and Non-Equilibrium Entropy,''
  Class.\ Quant.\ Grav.\  {\bf 27} (2010) 025006
  [arXiv:0906.4787 [hep-th]].


\bibitem{Bhattacharyya:2012ex}
  S.~Bhattacharyya,
  ``Constraints on the second order transport coefficients of an uncharged fluid,''
  arXiv:1201.4654 [hep-th].

\end{thebibliography}
\end{document}